\documentclass[twocolumn,aps,prc,showpacs,floatfix]{revtex4}
\usepackage{amsmath,bm}
\usepackage{graphicx}
\usepackage{epsfig}
\begin{document}

\title{Nuclear symmetry energy effects on liquid-gas phase transition in hot asymmetric
nuclear matter}
\author{Bharat K. Sharma and Subrata Pal}
\affiliation{Department of Nuclear and Atomic Physics, Tata Institute of 
Fundamental Research, Homi Bhabha Road, Mumbai 400005, India}

\begin{abstract}
The liquid-gas phase transition in hot asymmetric nuclear matter is investigated
within relativistic mean-field model using the density dependence of nuclear 
symmetry energy constrained from the measured neutron skin thickness of finite nuclei. 
We find symmetry energy has a significant influence on several features of 
liquid-gas phase transition. The boundary and area of the liquid-gas coexistence
region, the maximal isospin asymmetry and the critical values of pressure and 
isospin asymmetry all of which systematically increase with increasing softness 
in the density dependence of symmetry energy. The critical temperature below
which the liquid-gas mixed phase exists is found higher for a softer symmetry energy.

\end{abstract}

\maketitle
 
The possible occurrence of liquid-gas phase (LGP) transition in intermediate energy
heavy ion collisions using neutron-rich stable and future radioactive beams provides 
a rather unique tool to probe hot and dense phases of highly asymmetric nuclear matter.
Collisions experiments \cite{Poch,Elliot} with stable heavy nuclei at intermediate energy 
do indicate theoretically predicted \cite{Bondorf} features of liquid-gas phase 
transition where the hot and compressed nucleus produced expands and fragments 
into several intermediate mass fragments (high-density liquid phase) and 
light particles and nucleons (low-density gas phase).

The early theoretical studies of the thermodynamic properties of liquid-gas phase 
transition \cite{Lamb,Bertsch,Jaqaman,Lee} are mostly confined to symmetric nuclear 
matter that employed the quite well predicted \cite{Blaizot,Bruck,Siemens} behavior 
of the symmetric nuclear matter equation of state (EOS).
One of the major ingredients in studies of asymmetric nuclear matter require
knowledge of the density dependence of symmetry energy $E_{\rm sym}(\rho)$ 
\cite{Chen,BALi08,Baran}. Unfortunately, the model predictions of $E_{\rm sym}(\rho)$ 
even for nuclear matter at zero temperature are extremely diverse \cite{Danielewicz}. 
Only at the nuclear saturation density $\rho_0 \approx 0.16$ fm$^{-3}$ the value of 
$E(\rho_0, T=0) = 32\pm 4$ MeV has been well constrained.

Recently some progress has been achieved by consistently constraining
the symmetry energy of cold neutron-rich matter near normal matter density 
from analysis of isospin diffusion \cite{Chen,BALi08,Baran} and isoscaling 
\cite{Shetty} data in intermediate energy heavy ion collisions and from the 
study of neutron skin thickness of several nuclei \cite{Cent,Sharma}. While 
knowledge of symmetry energy $E_{\rm sym}(\rho,T)$ at finite temperature in 
particular has received little attention \cite{Dean,Xu,XuPLB} that is crucial 
for a proper understanding of the features of LGP transition in hot asymmetric 
nuclear matter. In fact, new qualitative features are expected when an asymmetric nuclear 
system with two conserved charges, baryon number and third component of isospin, 
undergoes a LGP change which has been suggested to be of second order \cite{Mueller}.
Most previous studies of LGP transition \cite{Mueller,Qian,Wang} relied on model 
predictions of symmetry energy $E_{\rm sym}(\rho,T)$ with no or minimal contact 
with the available experimental data. Thus to understand better the features of 
LGP transition in hot asymmetric nuclear matter, it is imperative to employ the 
asymmetric nuclear EOS that has been constrained from analysis of skin thickness 
data of several nuclei \cite{Sharma} or from isospin diffusion/scaling data 
\cite{BALi08,XuPLB}. Such an investigation is particularly useful as future 
experiments with radioactive ion beams with large neutron-proton asymmetries can 
be used to explore \cite{Brown,Horowitz} symmetry energy effects on 
liquid-gas phase transition.

In this letter, we study the effects of constrained symmetry energy \cite{Sharma} on the 
thermodynamic properties of LGP in hot neutron-rich nuclear matter within relativistic 
mean field (RMF) models \cite{Serot}. For this purpose we use two accurately calibrated models:
NL3 \cite{NL3} and FSUGold \cite{FSU}, that was obtained by fitting the model 
parameters to certain ground state properties of finite nuclei. The interaction 
Lagrangian density in the nonlinear RMF model is given by \cite{Horowitz,Sharma}
\begin{eqnarray} \label{RMF}
\mathcal{L} & = &\overline{\psi} \left[ g_s\phi
- \left(g_v V_\mu + \frac{g_\rho}{2} {\bm\tau} \cdot {\bf b_\mu}
+ \frac{e}{2}\left(1+\tau_3\right) A_\mu\right)\gamma^\mu \right]\psi \nonumber\\
& & -\frac{\kappa}{3!}\left(g_s\phi\right)^3
- \frac{\lambda}{4!}\left(g_s \phi\right)^4 
+ \frac{\zeta}{4!}g_v^4 \left( V_\mu V^\mu \right)^2 \nonumber\\
& & + \Lambda_v \left( g_\rho^2 {\bf b_\mu} \cdot {\bf b^\mu} \right) 
\left( g_v^2 V_\mu V^\mu \right)
\end{eqnarray}
which includes a isospin doublet nucleon field ($\psi$) interacting via exchange of 
isoscalar-scalar sigma ($\phi$), isoscalar-vector omega ($V^\mu$), isovector-vector 
rho ($\bf b^\mu$) meson fields and the photon ($A^\mu$) field. The nonlinear sigma 
meson couplings ($\kappa, \lambda$) soften the symmetric nuclear matter EOS 
at around $\rho_0$, while its high density part is softened by the 
self-interactions ($\zeta$) for the omega meson field. 

For the original NL3 set with $\zeta=\Lambda_v=0$, the saturation of
symmetric nuclear matter occurs at a Fermi momentum of $k_F = 1.30$ fm$^{-1}$ 
with a binding energy $B/A \approx 16.3$ MeV and an incompressibility of $K_0=271$ MeV. 
The original FSUGold \cite{FSU}, with two additional couplings $\zeta=0.06$ and 
$\Lambda_v=0.03$, with $K_0=230$ MeV produces a soft symmetric and asymmetric 
nuclear matter EOS. To study the effect of symmetric nuclear EOS 
(eg. incompressibility $K_0$) on the symmetry energy, we have extended 
\cite{Sharma} the original NL3 Lagrangian to include
the isovector coupling $\Lambda_v$ which is then varied along with $g_\rho$ 
in both NL3 and FSUGold to generate various $E_{\rm sym}(\rho)$.
All combinations of $\Lambda_v$ and $g_\rho$ are adjusted to a constant
$E_{\rm sym}(\overline\rho,T=0) = 25.67$ (26.00) 
for the NL3 (FSUGold) at an average density $\overline\rho$ corresponding to 
$k_F=1.15$ fm$^{-1}$ where the binding energy of $^{208}$Pb is reproduced. 
Thus the additional couplings provides an efficient way to change in a 
controlled manner the density dependence of nuclear symmetry energy without 
compromising the success of the model. 

The model parameter ($\Lambda_v$, $g_\rho$) set is then varied to explore 
$E_{\rm sym}(\rho)$ effects on the liquid-gas phase 
transition in hot asymmetric nuclear matter. For the present study we use
$\Lambda_v = 0.0 - 0.03$ since the resulting symmetry energies and their 
slopes and curvatures are in reasonable agreement with that extracted from 
neutron skin thickness of several nuclei as well as the isoscaling and isospin
diffusion data \cite{Sharma}. It may be also noted that with increasing $\Lambda_v$
the density dependence of symmetry energy becomes softer in both the 
NL3 and FSUGold models \cite{Sharma}. While at a finite $\Lambda_v$ 
the symmetry energy $E_{\rm sym}(\rho,T=0)$ is found to be particularly 
stiff in FSUGold than in the NL3 parameter sets at densities 
$\rho\gtrsim 1.5\rho_0$. 

At finite temperature and density the energy density ${\cal E}$ can be 
readily obtained from the thermodynamical potential $\Omega$ \cite{Mueller} as 
\begin{eqnarray} \label{eden}
{\cal E} & = & \frac{2}{(2\pi)^3} \sum_{q=n,p} \int d^3k \: E^*(k) 
\left( \left[n_q(k)\right]_+  + \left[n_q(k)\right]_- \right) \nonumber\\
&& + \frac{m_s^2 \phi^2}{2} + \frac{\kappa}{3!} \left(g_s\phi\right)^3
+ \frac{\lambda}{4!} \left(g_s \phi\right)^4  + \frac{m_v^2 V_0^2}{2} \nonumber\\
&& + \frac{\zeta}{8}\left(g_v V_0\right)^4 + \frac{m_\rho^2 b_0^2}{2} 
+ 3\Lambda_v\left(g_v V_0 \right)^2 \left(g_\rho b_0 \right)^2 ,
\end{eqnarray}
where $E^*(k) = \sqrt{k^2 + m^{* 2}}$ is the effective energy.
The distribution function for nucleon and antinucleon (referred to as $\pm$ sign)
\begin{equation} \label{DF}
\left[n_q(k)\right]_\pm = \frac{1}{\exp\left[\left(E^*(k)\mp \nu_q\right)/T\right]+1}
~~~~ (q = n,p) ,
\end{equation}
where the effective chemical potential for neutron and proton is expressed as  
$\nu_q = \mu_q - g_v V_0 \pm g_\rho b_0/2$. The chemical potentials can be determined 
from the conserved baryon and isospin densities:
\begin{eqnarray} \label{dens}
\rho = \frac{2}{(2\pi)^3} \int d^3k \: (G_p(k) + G_n(k)) ,\\ 
\rho_3 = \frac{2}{(2\pi)^3} \int d^3k \: (G_p(k) - G_n(k)) ,
\end{eqnarray}
where $G_q(k) = \left[n_q(k)\right]_+ - \left[n_q(k)\right]_-$.

\begin{figure}[ht]
\vspace{.2cm}

\centerline{\epsfig{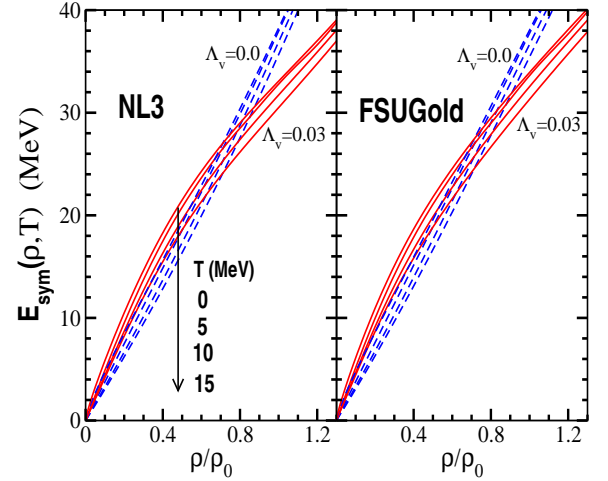}}
\caption{Density dependence of nuclear symmetry energy at temperatures
$T=0,5,10,15$ MeV in the NL3 (left panel) and FSUGold set (right panel)
with couplings $\Lambda_v = 0.0$ and 0.03.} 

\label{esym}
\end{figure}

As in the zero temperature case, several model studies \cite{BALi08,Xu,Zuo,Mosh} 
have indicated that the EOS for hot neutron-rich nuclear matter can be expressed in
the parabolic form:
\begin{equation} \label{EOST}
E(\rho,T,\alpha) = E(\rho,T,\alpha=0) + E_{\rm sym}(\rho,T)\alpha^2 
+ {\cal O}(\alpha^4) ,
\end{equation}
where the neutron-proton asymmetry is $\alpha = (\rho_n-\rho_p)/\rho$. 
The density and temperature dependence of symmetry energy can be estimated
from $E_{\rm sym}(\rho,T) \approx E(\rho,T,\alpha=1) - E(\rho,T,\alpha=0)$.
This implies that $E_{\rm sym}(\rho,T)$ is the energy required to convert 
all the protons in symmetric matter to neutrons. Figure \ref{esym} shows
the density dependence of nuclear symmetry energy at temperatures $T=0,5,10,15$ MeV
in the NL3 (left panel) and FSUGold (right panel) sets. 
For all choices of $\Lambda_v$ the symmetry energy decreases with increasing 
temperature especially at small densities $\rho \lesssim \rho_0$ that is entirely 
due to the decrease in the kinetic energy contribution.
For $\Lambda_v=0.0$ (0.03) the density dependence of $E_{\rm sym}(\rho,T)$ at
all temperatures exhibits a systematic trend of small (large) value at 
subsaturation densities and a large (small) value at supranormal densities 
resulting in an overall stiffer (softer) asymmetric nuclear matter EOS.

\begin{figure}[ht]
\vspace{.2cm}

\centerline{\epsfig{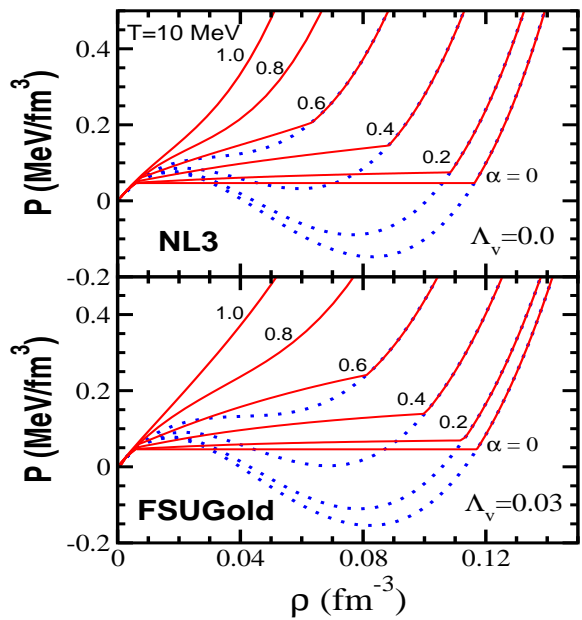}}
\caption{Pressure as a function of density at temperature $T = 10$ MeV for 
various isospin asymmetry $\alpha$ in the original NL3 set \cite{NL3} with
$\Lambda_v=0.0$ (top panel) and the original FSUGold set \cite{FSU}
with $\Lambda_v = 0.03$ (right panel). The dotted curves refer to unstable
single phase while the solid curves refer to stable matter; see text for details.}

\label{eos}
\end{figure}

The above described models can now be used to study LGP in hot asymmetric nuclear 
matter. The system is stable against LGP separation if its free energy 
$F$ is lower than the coexisting liquid ($L$) and gas ($G$) phases, i.e. 
$F(T,\rho) < (1-\lambda) F^L(T,\rho^L) + \lambda F^G(T,\rho^G)$ with
$\rho = (1-\lambda) \rho^L + \lambda \rho^G$ where $0< \lambda <1$ and 
$\lambda = V^G/V$ being the fraction of the total volume occupied by 
the gas phase. The stability condition implies the inequalities \cite{Mueller}:
\begin{eqnarray} \label{MEC}
\rho\left(\frac{\partial P}{\partial \rho}\right)_{T,\alpha} > 0 ,
\end{eqnarray}
\begin{eqnarray} \label{CHM}
\left(\frac{\partial \mu_{p}}{\partial \alpha}\right)_{T, P} < 0
~~~ {\rm or} ~~~
\left(\frac{\partial \mu_{n}}{\partial \alpha}\right)_{T, P} > 0 .
\end{eqnarray}
The first inequality indicates mechanical stability which means a system
at positive isothermal compressibility remains stable at all densities. The second
inequality stems from chemical instability which shows that energy is required
to change the concentration in a stable system while maintaining temperature and pressure
fixed. If one of these conditions get violated, a system with two phases is
energetically favorable. 
The two phase coexistence is governed by the Gibbs's criteria for equal chemical 
potentials and pressures in the two phases with different densities but at the 
same temperature:
\begin{eqnarray} 
\mu_q^L(T,\rho^L) &=& \mu_q^G(T,\rho^G) ~~~~ (q = n,p) , \label{GBS1} \\
P^L(T, \rho^L) &=& P^G(T,\rho^G) .  \label{GBS2}
\end{eqnarray}

\begin{figure}[ht]
\vspace{.2cm}

\centerline{\epsfig{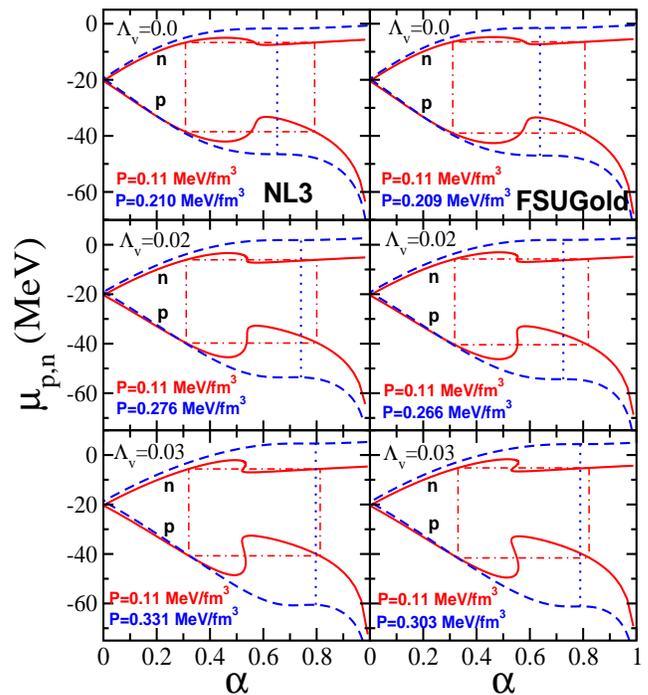}}
\caption{Chemical potential isobars as a function of isospin asymmetry $\alpha$ 
at temperature $T = 10$ MeV for NL3 (left panel) and FSUGold (right panel) with 
different $\Lambda_{v}$ couplings. The geometrical construction used to obtain 
the isospin asymmetries and chemical potentials in the two coexisting phases 
is also shown.}

\label{chem}
\end{figure}

Figure \ref{eos} shows the pressure as a function of nucleon density at a 
fixed temperature $T$ = 10 MeV with different values of asymmetry $\alpha$ in the
original NL3 and FSUGold sets. Below a critical value of asymmetry $\alpha$, 
the pressure is seen (dotted curves) to decrease with 
density resulting in negative incompressibility and thereby a mechanically 
unstable system. The stable two-phase (liquid-gas) configuration at each density
is obtained from Maxwell construction (solid lines). Analogues to intermediate
energy heavy-ion collisions \cite{Poch,Elliot} when the hot matter in the 
high density (liquid) phase expands it enters the coexistence LGP where the 
pressure decreases at a fixed $\alpha \neq 0$ for the two-component asymmetric 
matter. Whereas, for symmetric nuclear matter at $\alpha = 0$ the pressure 
remains constant at all densities. Finally the system leaves the coexistence 
region and vaporizes into the low density (gas) phase. Of particular interest here
is the symmetry energy effects on the isotherms. It is clearly seen that in
contrast to the original NL3 with $\Lambda_v=0$, the softer $E_{\rm sym}(\rho)$ 
in the original FSUGold with $\Lambda_v=0.03$ \cite{Sharma} enforces the
onset of pure liquid phase to a higher density resulting in a wider
coexistence region for each asymmetry $\alpha$. Moreover, the critical
pressure $P_c$ above which the mixed liquid-gas phase vanishes is seen larger
for this soft FSUGold set; a detailed discussion of which is presented below.

The details of chemical evolution for the LGP transition is depicted in Fig.
\ref{chem} where the neutron and proton chemical potentials are shown
as a function of isospin asymmetry $\alpha$ at a fixed $T=10$ MeV and pressure
$P=0.11$ MeV/fm$^3$ for the NL3 (left panels) and FSUGold (right panels) at
various $\Lambda_v$ values. As usual, the bare nucleon mass has been subtracted
from the chemical potentials.
At fixed pressure and $\Lambda_v$, the solutions of the Gibbs conditions 
(\ref{GBS1}) and (\ref{GBS2}) for phase equilibrium form the edges of a rectangle
and can be found by geometrical construction as shown in Fig. \ref{chem}.
At each $\Lambda_v$, the two different values of $\alpha$ defines the high
density liquid phase boundary (with  small $\alpha = \alpha_1(T,P)$) and the
low density gas phase boundary (with large $\alpha = \alpha_2(T,P)$). From the
figure it is evident that the symmetry energy dependence of $\Lambda_v$ in
NL3 and FSUGold \cite{Sharma,Horowitz} leads to different phase boundaries 
$\alpha_1(T,P)$ and $\alpha_2(T,P)$ and hence should predict different 
thermodynamic properties for the LGP transition. 

As the pressure increases the system encounters a critical pressure $P_c$ 
beyond which the matter is stable but below which the second inequality 
(\ref{CHM}) gets violated and the system becomes chemically unstable. The 
critical pressure $P_c$ is determined by the inflection 
point $(\partial \mu/\partial\alpha)_{T,P_c} 
= (\partial^2 \mu/\partial\alpha^2)_{T,P_c} = 0$. The disappearance of chemical
instability at $P_c$ results in the neutron (proton) chemical potential to
decrease (increase) with decreasing asymmetry $\alpha$. Figure \ref{chem} also 
shows the chemical potential isobars at the critical pressure (dashed lines). The rectangle
from Gibbs condition then collapses into a line vertical at $\alpha \equiv \alpha_c$.
Correspondingly, ($P_c,\alpha_c$) defines the critical point at a given temperature
that refers  to the upper boundary of instability with respect to pressure 
variation. Note at $T=10$ MeV, the critical values ($P_c,\alpha_c$) 
at $\Lambda_v=0.0, 0.02, 0.03$ are respectively at
(0.210, 0.652), (0.276, 0.741), (0.331, 0.797) for the NL3 set and at
(0.209, 0.638), (0.266, 0.725), (0.303, 0.789) for the FSUGold set.
Interestingly, we also find at a finite temperature the stiffness of symmetry energy
has a significant influence on the phase-separation boundaries of LGP transition 
\cite{XuPLB}. In general, a softer symmetry energy (larger $\Lambda_v$) 
gives systematically larger 
critical pressure and an enhanced asymmetry in the system. Moreover at a 
finite $\alpha$, the relatively softer symmetry energy $E_{\rm sym}(\rho, T=0)$ 
in the NL3 compared to FSUGold \cite{Sharma} translates to a larger 
critical pressures and asymmetry for the LGP transition.

\begin{figure}[ht]
\vspace{.2cm}

\centerline{\epsfig{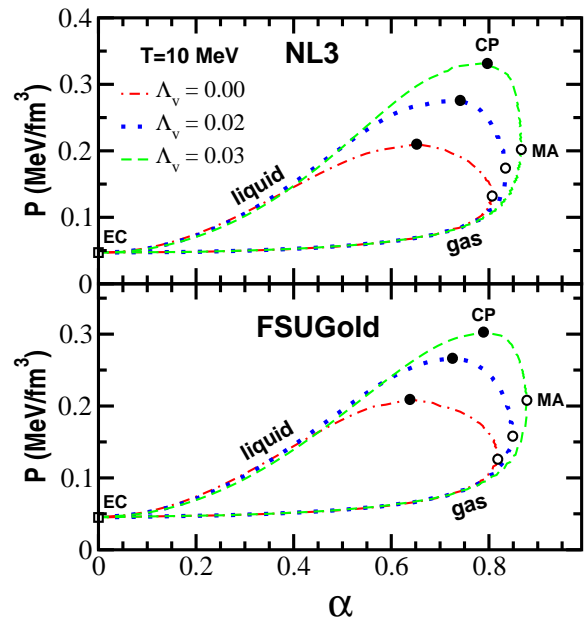}}
\caption{The section of binodal surface at temperature $T$ = 10 MeV in NL3 
(top panel) and FSUGold (bottom panel) with different $\Lambda_v$ couplings.
The critical point (CP), the point of equal concentration (EC), and the maximal
asymmetry (MA) are indicated.}

\label{binod}
\end{figure}

All the pairs of solutions of Gibbs conditions, $\alpha_1(T,P)$ and $\alpha_2(T,P)$, 
form the phase-separation boundary or the binodal surface. In Fig. \ref{binod} 
we show the section of the binodal surface under isothermal compression of 
asymmetric nuclear matter at $T=10$ MeV in the NL3 (top panel) and FSUGold 
(bottom panel). As expected the point of equal concentration (EC) corresponding 
to symmetric nuclear matter is independent of $\Lambda_v$. The critical point
(CP) and EC divide  the binodal section into two branches. One branch is the high-density
(liquid) phase that is less asymmetric while the other branch corresponds to the
more asymmetric low-density (gas) phase. Thus the matter on the left (right) of the
binodal surface represents stable liquid (gas) phase. It is clearly seen here that 
the critical point ($P_c, \alpha_c$) depends on the density dependence of the symmetry
energy associated with different $\Lambda_v$ values.

We also indicate on the binodal surface the maximal isospin asymmetry (MA), $\alpha_{\rm MA}$,
of the system.  Thus more neutron-rich matter on the right side of the surface when 
compressed/expanded at fixed $\alpha$ will never encounter a coexistence phase. 
Note here the maximal asymmetry is also quite sensitive to $\Lambda_v$ i.e. 
on $E_{\rm sym}(\rho,T)$. Such effects found in the present study should have 
strong influence on the experimentally observed isospin distillation phenomena
\cite{XuISF} where the gas phase is more neutron-rich (large $n/p$ ratio)
compared to the more asymmetric liquid phase. However for pressures 
$P \geq 0.10$ MeV/fm$^3$ the magnitude of isospin distillation is more sensitive
to the symmetry energy used. 

A new feature for LGP transition in asymmetric system, refereed to as retrograde 
condensation \cite{Mueller}, arises when a nucleon gas prepared at an 
asymmetry $\alpha_c < \alpha < \alpha_{\rm MA}$ is compressed
at fixed total $\alpha$. The matter remains mechanically stable but chemically 
unstable. Thus a coexisting liquid phase emerges which finally vanishes when 
the system leaves the binodal surface as a pure gas.
As the extent $\Delta\alpha = \alpha_{\rm MA} - \alpha_c$
is found to decrease for softer symmetry energy with higher $\Lambda_v$, the 
possibility of such unique-phase condensation phenomena also becomes minimal.

The present study clearly suggests that for liquid-gas phase transition in hot
asymmetric nuclear matter, the critical values of pressure and isospin asymmetry, the
maximal asymmetry and the area and shape of the binodal surface are quite sensitive
to the density dependence of symmetry energy with a stiffer symmetry energy leads
to consistently smaller values of these thermodynamic variables.

\begin{figure}[ht]
\vspace{.2cm}

\centerline{\epsfig{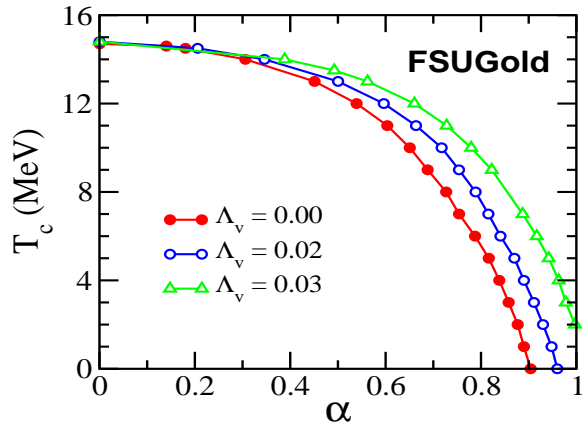}}
\caption{The critical temperature $T_c$ versus isospin asymmetry $\alpha$ for 
different $\Lambda_v$ in the FSUGold set.}

\label{Talpha}
\end{figure}

The existence of critical isospin asymmetry parameter $\alpha_c$ at a given temperature
indicates that for $\alpha > \alpha_c$ the system will not change completely into the
liquid phase. Conversely, this suggests that at a fixed $\alpha$ there exists a critical
temperature $T_c$ beyond which the system can only be in the gas phase at all pressures.
In Fig. \ref{Talpha} we present $T_c$ as a function of $\alpha$ in the FSUGold
set for different couplings $\Lambda_v$. For symmetric nuclear matter ($\alpha=0$),
the critical temperature for LGP transition in this model is $T_c = 14.7$ MeV.
With increasing asymmetry $\alpha \gtrsim 0.6$, $T_c$ decreases rapidly.
A softer density dependence in symmetry energy (larger $\Lambda_v$) shows the
coexisting liquid-gas phase can prevail for larger values of $T_c$. 
We find that for the soft symmetry energy ($\Lambda_v=0.03$) even pure
neutron matter ($\alpha=1$) can exhibit LGP transition at $T \leq T_c = 2$ MeV.
While the stiffest symmetry energy ($\Lambda_v=0$) at $\alpha > 0.9$ predicts
that the matter can only be in the pure gas phase at all temperatures. 

In summary the effects of isospin symmetry interaction on the liquid-gas phase 
transition in hot neutron-rich nuclear matter is investigated. For this we have
used the two accurately calibrated relativistic mean field models, the NL3 
\cite{NL3} and the FSUGold \cite{FSU} wherein the density dependence of nuclear 
symmetry energy at zero temperature has been constrained within a limited range 
by neutron skin thickness data of several atomic nuclei. We find considerable
sensitivity of the symmetry energy on the features of phase transition.
Softer symmetry energies give progressively larger phase-separation boundaries 
with higher critical values for pressure and isospin asymmetry as well as
maximal asymmetries. At a given asymmetry we find the critical temperature 
for the existence of the mixed liquid-gas phase increases with softer symmetry
energy and predicts the possible occurrence of even an unstable pure neutron 
matter at finite temperatures.

\end{document}